\documentclass{article}
\usepackage[utf8]{inputenc}
\usepackage{amsmath}
\usepackage{fouriernc}
\usepackage{gensymb}
\usepackage{siunitx}
\usepackage{float}
\usepackage{authblk}
\usepackage{verbatim}
\usepackage{tikz}
\usepackage{pgfplots}
\usetikzlibrary{calc}
\usetikzlibrary{arrows}
\usepackage{graphicx}
\usepackage{subcaption}
\usepackage{hyperref}
\hypersetup{colorlinks=true,allcolors=blue}
\usepackage[authoryear]{natbib}
\setcitestyle{authoryear,open={(},close={)}}
\pgfplotsset{width=8cm, compat=1.9}

\title{\LARGE\textbf{On the investigation of two non-neutral static bodies}}
\author{Roshan Raj}
\date{May 09, 2022}
\affil{\small kaalkrit@outlook.com\\Int. M.Sc.-III\\Dept. of Physics, SVNIT Surat }
\begin{document}
\maketitle

\begin{abstract}
    In this work, we have considered statics of two non-neutral bodies unaffected by external force fields. We have tried to calculate the value of $\theta$ at which net field strength becomes zero however under given boundary condition, we get no solution. Thereafter, We define a function $Y(A,\theta)$, it can be found that the function $Y$ mathematically resembles to Semi-mass empirical function $M(Z,A)$ that exists in nuclear model. Further investigations on $Y$ has shown that \textit{the nature gives double preference to a system ,  which is having heavier object at center of positive charge over another negative charged body}. Meanwhile, we come across another expression which depicts about the geometry of ellipse in complex plane under certain conditions, Whose physical significance is yet to be get realized. Lastly, we have investigated a case associated with non-zero net field. As a result of it, we define a new quantity as \textit{quintessence}, which behaves analogous to the electric potential, and it is mathematically turned out  to be a generating function for Legendre polynomials, which also highlights the previous statement about nature's selective behaviour. We have tried to verify the constancy of a factor $\chi$ which runs throughout the various expressions in this study of statics, it is indeed found to be almost constant but highly sensitive to scale or order of size of the subject under study.
\end{abstract}

\section{Introduction}
The following work,  we have assumed $G_\pm$  of density $d_g$ and $J_+$  of density $d_j$ as the tiny rigid charged body \footnotemark{\footnotetext{You may have noticed that i did not specify the kind of charge body $G_\pm$ posses, thus it is being sub-scripted  by $\pm$  symbol, depicting that either positive or negative charged G. Later it will take us to different sub cases. }} and the spherical fluid bulk of positive charge respectively, where $d_g >> d_j$ .\\
In particular, we will be discussing in detail about the statics of two non neutral bodies  $G_\pm$ and, $J_+$. We will be considering two cases about the body $G_\pm$ signifying that it might be either positive or negative while larger bulk should always be positive. 
In the first subsection (2.1.1) we have tried to find out spatial point(s) having absence of both gravitational and electric fields, which exists either due to balancing of forces or repulsion of like charges. In the second subsection (2.1.2) We will coming across to a striking mathematical similarity between semi-mass empirical formula (SMEF) of odd A nucleus, with our own expression of $Y(A,\theta)$, Which have tempted us to coin a term \textit{specific field parabolas} in analogous to \textit{mass parabolas}. In the third subsection (2.1.3), we come to know that nature prefers a system comprised of two unlike charges such that positive one stay at the centre and negative surrounds the former over to that system which is having negative charged centre and positive surroundings. Lastly, in the fourth subsection (2.1.4), we have defined a quantity named as \textit{quintessence }, having inverse proportionality to net field strength, which encapsulates the collective meaning of both mass and electric charge. We see that this quantity can be expanded in similar fashion as we do for multi pole expansion of electric potential. We also try to verify our one of the assumptions that we will be making in regard the constancy of factor $\chi$. 
 
\section{Formulation }
\label{section2}
We understand that the tiny body $G_\pm$ if carries electric charge  of magnitude $Q_g$ and having mass $M_g$ will only be acted by net electric and gravitational forces. Let stationary bulky body $J_+$ has electric charge of magnitude $Q_j$ and mass $M_j$. In order to apply both Coulomb's electrostatics and Newton's gravitational expressions, we assume that net value of both the total charges and mass are being concentrated at the geometric centre of each body, when these bodies are at rest with respect to each other. Otherwise, In the case of motion, we will need to consider for infinitesimal elements of charge and mass, and their contributions that would be producing net fields. It should be noted that we are considering our system as an ideal one.
\subsection{Static System}
\subsubsection{Searching for null points}
\label{s1}

Suppose there exist points $B_n$ in space, where vector summation of all the force fields from each body become zero.  Let $\theta_g$ and $\theta_j$ are the respective angles made with axial line ($x'x$)  by the lines joining the point $B_n$, (where $n$ is any integer number) to the respective centres of bodies $G_\pm$ and $J_+$ of distance $l_g$ and $l_j$.
\begin{figure}[ht]
    \centering
     \includegraphics{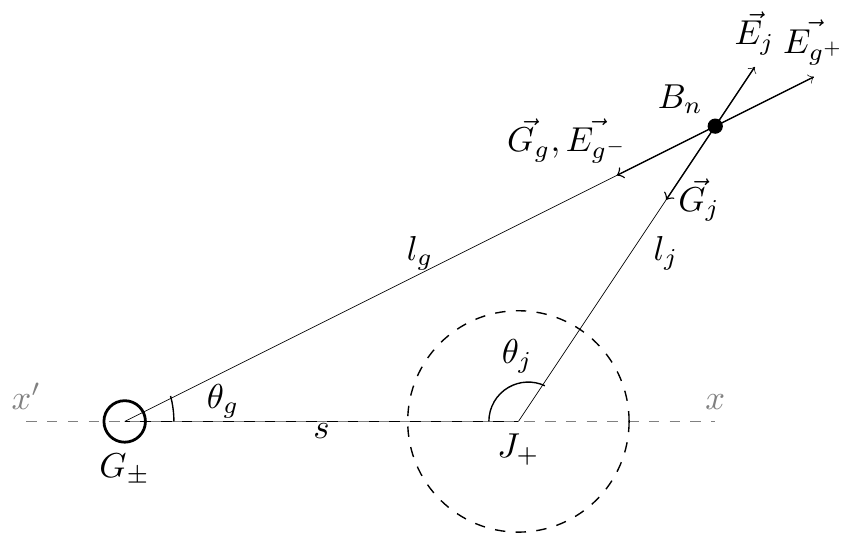}
    \caption{ \small Null point at $B_n$. Here, vector summation of all the force fields from each body becomes zero.}
    \label{fig:1}
\end{figure}
Here, We deduce the net electric field strength at any point $B_n$ due to both the bodies as
$\vec{E}=\vec{E_g}+\vec{E_j}$
where, \begin{equation}
\vec{E_g}=\pm E_g \cos(\theta_g)\hat{x}\pm E_g \sin(\theta_g)\hat{y}
\nonumber
\end{equation}
\begin{equation}
\vec{E_j}= -E_j \cos(\theta_j)\hat{x}+ E_j\sin(\theta_j)\hat{y}
\nonumber
\end{equation}

similarly, 
the net gravitational field strength at any point $B_n$ be
$\vec{G}=\vec{G_g}+\vec{G_j}$
where, 
\begin{equation}
\vec{G_g}=-G_g \cos(\theta_g)\hat{x}- G_g \sin(\theta_g)\hat{y}
\nonumber
\end{equation}
\begin{equation}
\vec{G_j}= + G_j \cos(\theta_j)\hat{x}- G_j\sin(\theta_j)\hat{y}
\nonumber
\end{equation}
now net field ($\vec{f} $) along x-direction  and y-direction respectively be 

 \begin{equation}
 \vec{f_x}=\left(\pm E_g -\chi G_g\right) \cos(\theta_g)\hat{x}+(- E_j  + \chi G_j) \cos(\theta_j)\hat{x}
 \nonumber
 \end{equation}
 \begin{equation}
 \vec{f_y}=\left(\pm E_g - \chi G_g\right) \sin(\theta_g)\hat{y}- (-E_j+ \chi G_j)\sin(\theta_j)\hat{y}
 \nonumber
 \end{equation}

now, magnitude of net field be \footnotemark{\footnotetext{We can reduce the equation (\ref{e2.1.1}) into $E_{net}=\frac{sq}{4\pi\epsilon_0 r^3}\sqrt{3\cos{\Theta}^2 +1}$, if we ignore the higher powers of $\frac{s}{2}\cos{\Theta}$ along with any gravitational effects and using relations $l_g=r-\frac{s}{2}\cos{\Theta}$ , $l_j=r+\frac{s}{2}\cos{\Theta}$  and $\Theta=2\pi n \pm\cos^{-1}\left({\pm \frac{\sqrt{\cos{\theta}}}{\sqrt{3}}}\right)$. Also taking, $Q_g$ and $Q_j$ as respective negative and positive charges of equal magnitude, $q$ and $n$ is the element in $\mathbb{Z}$.}}, $|\vec{f}|=\sqrt{|\vec{f_x}|^2+|\vec{f_y}|^2}$,

\begin{equation}
|\vec{f}|=\left[{2(\pm E_g -\chi G_g)(-E_j+ \chi G_j)\cos(\theta)+(\pm E_g - \chi G_g)^2+(-E_j+ \chi G_j)^2}\right]^{1/2}
\label{e2.1.1}
\end{equation}

Where, $(\theta=\theta_g+\theta_j)$, $\chi$ is a factor having physical dimension of $[M T^{-1} A^{-1}]$ to make the expression dimensionally consistent  and just refer it as an Chi coefficient and $G_{g}=\frac{G_NM_{g}}{l_g^2}$, $E_g=\frac{1}{4\pi\epsilon_0}\frac{Q_g}{l_g^2} $ and so on for $J_+$. Here, we have assumed that $\chi$ is independent of mass and charge, that's why we have considered this factor equivalent for both the bodies $G_\pm$ and $J_+$ .

Rearranging the above expression and factoring out $(\chi G_j-E_j)$ from equation (\ref{e2.1.1}), we get,
\begin{equation}
|\vec{f}|=(\chi G_j-E_j)\sqrt{1-2\left(\frac{\chi G_g\mp E_g}{\chi G_j-E_j}\right)\cos(\theta)+\left(\frac{ \chi G_g\mp E_g }{\chi G_j-E_j}\right)^2}
\label{e2.1.2}
\end{equation}
If the magnitude of net field at any null point $B_n$ will be zero, then we get
\begin{equation}
0=1-2\left(\frac{\chi G_g\mp E_g}{\chi G_j-E_j}\right)\cos(\theta)+\left(\frac{ \chi G_g\mp E_g }{\chi G_j-E_j}\right)^2
\nonumber
\end{equation}
note here, we have accepted that $(\chi G_j-E_j)\neq 0 $  otherwise we could not have gone further to find out null points. Additionally, we also see, $\frac{|\chi G_g\mp E_g|}{|\chi G_j-E_j|}<1$, and substituting the values of gravitational and electric field strength due to $G_\pm$ and $J_+$ bodies, in order to find null points. 

\begin{equation}
0=1-2\frac{l_j^2}{l_g^2}\left(\frac{\chi G_NM_{g}\mp \frac{1}{4\pi\epsilon_0}Q_g}{\chi G_NM_{j}-\frac{1}{4\pi\epsilon_0}Q_j}\right)\cos(\theta)+\left(\frac{l_j^2}{l_g^2}\right)^2\left(\frac{\chi G_NM_{g}\mp \frac{1}{4\pi\epsilon_0}Q_g}{\chi G_NM_{j}-\frac{1}{4\pi\epsilon_0}Q_j}\right)^2
\nonumber
\end{equation}
Or,
\begin{equation}
A^2\left(\frac{l_j^2}{l_g^2}\right)^2-2A\cos(\theta)\left(\frac{l_j^2}{l_g^2}\right)+1=0
\label{e2.1.3}
\end{equation}

where, $\frac{1}{A}=\left(\frac{\chi G_NM_{j}-\frac{1}{4\pi\epsilon_0}Q_j}{\chi G_NM_{g}\mp \frac{1}{4\pi\epsilon_0}Q_g}\right)=\frac{M_j}{M_g}\left(\frac{1-\frac{\gamma}{\chi }\frac{Q_j}{M_j}}{1 \mp \frac{\gamma}{\chi}\frac{Q_g}{M_g}}\right)$; Here, $\gamma=\frac{1}{4\pi\epsilon_0 G_N}\approx 1.346\times10^{20}$ $\frac{\text{kg}^2}{\text{C}^2}$ and $\chi$ is unknown constant.
If we solve for $\theta$ from above expression, if $\frac{l_j}{l_g}A\neq 0$, along with $|A|< 1$, we simply get, 

\begin{equation}
  \theta=B_n=2\pi n \pm \cos^{-1}\left[{\frac{(\omega A)^2 +1}{2\omega A}}\right]
  \label{e2.1.4}
\end{equation}
Where $n$ is the element in $\mathbb{Z}$ and $\omega=\left(\frac{l_j}{l_g}\right)^2$. from above relation we get that, null points $B_n=B_n(\omega,A)$. From above we see that $\omega>0$ along with restricting  $0<A<1$ can only give us meaning values.\footnotemark{\footnotetext{We will see later, why positive $A$ is essential to assume here only, see section (\ref{section2.1.3}).}}

In particular, if we compel for a moment to make $\left(\frac{a^2+1}{2a}\right)< 1$,  also encapsulating $\omega A=a$, such that $0<a<1$.
We ultimately see that for any value of $a$ under given boundary condition, we do not get any non-trivial solution. However, \textit{iff $a\rightarrow1$} we do see (equation (\ref{e2.1.3}))  an integer solution giving $\theta \rightarrow 0$ and \textit{then $B_0=0$}.
Therefore, $\theta_g=-\theta_j$ or 
\begin{equation}
   |\theta_g|=|\theta_j| 
   \label{s1r}
\end{equation} It should only give the plane containing full of null points, those running normal to the (\textit{x'x}) line. Consequently, we note that both the side lengths should be equal in their magnitudes.

\subsubsection{Drawing  mathematical resemblances with mass parabolas }
\label{s2}
The following treatment made over expression (\ref{e2.1.3}) has came to me as a surprise, irrespective of fact that any two quadratic expressions can be made to appear similar  mathematically (parabolic in form) but the following correspondences are shown seem definitely intriguing . Since, I was not expecting such result. It was found to be analogous to expressions, we generally get while striving to find most stable isobar in nuclear model through semi-empirical mass formula. 
We can rewrite the equation (\ref{e2.1.3}) as \begin{equation}
    Y(A,\theta)=A^2\omega^2 -2A\omega cos(\theta)+1
    \label{e2.1.2.1}
\end{equation}
 Now, partially differentiate and equate it to zero, \begin{equation}
     \left(\frac{\partial Y}{\partial A}\right)_{\theta=const.} =2A_0\omega^2 -2\omega\cos(\theta) =0
 \nonumber
 \end{equation}
giving $A_0$ = minima of function $Y(A,\theta)$ at constant\footnotemark{\footnotetext{Remember that $\theta=\theta_g +\theta_j$ , if we had considered $\theta$ to be a constant, then we could still vary up the magnitudes of both $\theta_g $  and $\theta_j$ such that the $\theta$ could be remaining unaltered. Also understand the reason of why we did not consider Y as function of $\omega$  also but $Y(A,\theta)$, was due to the fact that both $\omega=\left(\frac{l_j}{l_g}\right)^2=\omega(\theta_g, \theta_j)=\left|\frac{\sin(\theta_g)}{\sin(\theta_j)}\right|^2  $after applying law of sine, (see figure-\ref{fig:1} ) and $ \theta=\theta(\theta_g, \theta_j)$ as mentioned above.}} magnitude of  $\theta$.
\begin{equation}
    A_0=\frac{\cos(\theta)}{\omega}
    \label{e2.1.2.2}
\end{equation}
magnitude of function be
\begin{equation}
\begin{split}
Y(A_0,\theta) &=\left(\frac{\cos(\theta)}{\omega}\right)^2\omega^2
-2\left(\frac{\cos(\theta)}{\omega}\right)\omega cos(\theta)+1\\
&=\sin^2(\theta)\\
&=1-A_0^2\omega^2
\label{e2.1.2.3}
\end{split}
\end{equation}
now, we also see that
\begin{equation}
\begin{split}
    Y(A,\theta)-Y(A_0,\theta) & =A^2\omega^2 -2A\omega cos(\theta)+1-1+A_0^2\omega^2\\
  & =A^2\omega^2-2AA_0\omega^2+A_0^2\omega^2\\
  & =\omega^2(A-A_0)^2
\end{split}
\label{e2.1.2.4}
\end{equation}

\begin{figure}
    \centering
    \includegraphics[width=0.6\textwidth]{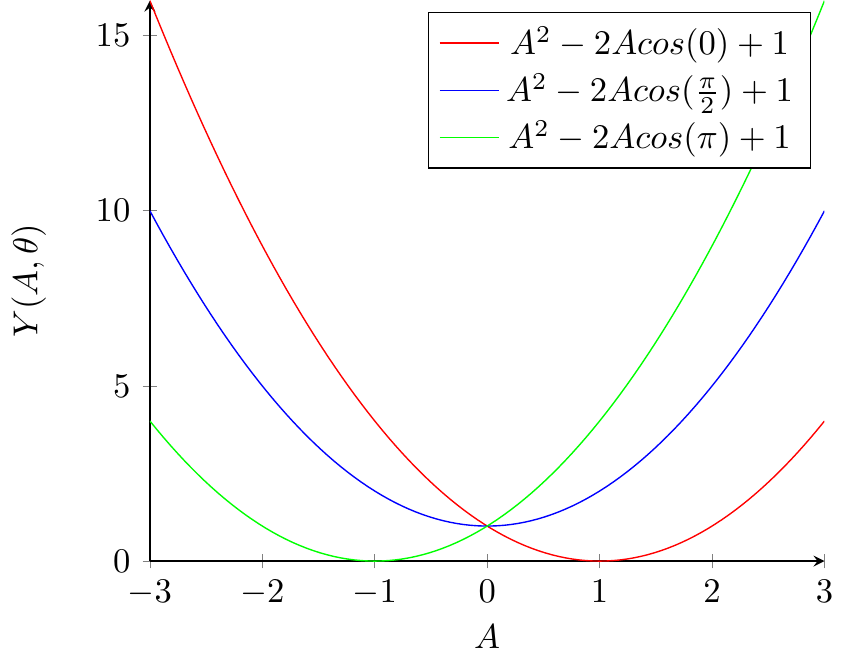}
    \caption{ \small Different specific field parabolas having $\omega=1$. }
    \label{fig-para}
\end{figure}

In the this paragraph, the notation we would be considering are not equivalent to prior notations we have been using so far, but only particularly here, A is atomic mass number, Z is atomic number and symbols such as $\alpha$, $\beta$ \& $\gamma$ are coefficient having dependency on $A$.( see \cite{patel1991nuclear})
Now, we seek for analogies from semi-mass empirical formula (SMEF) of odd $A$ nucleus\footnotemark{\footnotetext{Sometimes also called the Weizsäcker formula, Bethe–Weizsäcker formula, or Bethe–Weizsäcker mass formula}}, which is expressed as $ M(Z,A) = \alpha A+\beta Z+\gamma Z^2 $, when solved for nuclear charge of most stable isobar gives $ Z_0 = -\frac{\beta}{2\gamma}$. Ultimately, $ M(Z_0,A) = \alpha A-\gamma Z_0^2 $ and difference of masses, which are getting equal to $\gamma(Z-Z_0)^2$ are respectively sharing mathematical resemblance with above equations from (\ref{e2.1.2.1} to \ref{e2.1.2.4}). Where SMEF parameters (left) seem analogous\footnotemark{\footnotetext{Analogous in two sense, firstly through mathematical form and secondly in physical dimension wise, see both Z and A are dimensionless.}} to our variables (right), such as $Z \cong A $, $A \cong \theta $, $\gamma \cong \omega^2 $, $\beta \cong -2\omega \cos(\theta)$, $\alpha\cong\frac{1}{A}$ and then obviously, $M(Z,A)\cong Y(A,\theta)$.\\
We define $Q_{\theta} =Y(A_1,\theta)-Y(A_2,\theta)$ for any two values of 
$A\neq 0$, it gives,
\begin{equation}
Q_{\theta}=\omega^2(A_1 +A_2 -2A_0)(A_1-A_2)
\label{e2.1.2.5}
\end{equation}
Which find its resemblance with $Q_{\beta^\pm}$ values in beta decay.
We know that expression $M(Z,A)$ if it were to be plotted against $Z$ then it would have displaying a curve generally called as \textit{mass parabolas}, similar trends we expect to observe by the plot of equation (\ref{e2.1.2.1}) against $A$ with $A_0$ giving the minima of the curve,(see figure \ref{fig-para}). Let's call that as \textit{specific field parabolas}.
In nuclear physics, central nucleus of an atom modeled as a liquid drop was proposed by George Gamow \citep{Gamow1930}, similarly in our case we considered $J_+$ as positively charged fluid bulk, however we did not consider any factor generally associated to a fluid nature of the object. We hope that while doing dynamics, it may help us to interpret the above resemblances. 

\subsubsection{Natural preference given to the atom like system }\label{section2.1.3}

Also looking carefully to equation (\ref{e2.1.3}) or function $Y(A,\theta)=0$ again, we see a quadratic equation in $\left(\frac{l_j}{l_g}\right)^2$, which has two roots as,
\begin{equation}
\frac{l_j^2}{l_g^2}=\frac{\cos(\theta)\pm i\sin(\theta)}{A}=\frac{e^{\pm i\theta}}{A}
\nonumber
\end{equation}
Or,
\begin{equation}
\frac{l_j}{l_g}=\frac{1}{\sqrt{A}}e^{\pm\frac{ i \theta}{2}}
\label{e2.1.5}
\end{equation} We have neglected the negative square root as ratio of two lengths cannot be a negative number. Now using, law of sine, we get $\frac{l_j}{l_g}=\left|\frac{\sin (\theta_g)}{ \sin(\theta_j)}\right|$ and product law of exponent, $e^{\pm\frac{ i \theta}{2}} = e^{\pm\frac{ i \theta_g}{2}}\cdot e^{\pm\frac{ i \theta_j}{2}}$ and rearranging the expression. gives,
\begin{equation}
|\sin (\theta_g)|\cdot e^{\mp\frac{ i \theta_g}{2}}=\frac{1}{\sqrt{A}}e^{\pm\frac{ i \theta_j}{2}}\cdot |\sin(\theta_j)|
\nonumber
\end{equation}
Now,we bring left side of expression to the right side, giving us all the expression equal to zero. We again define a function, say $ C(\theta_g, \theta_j)$ that encapsulates all those right terms in itself.
We have,
\begin{equation}
    C(\theta_g, \theta_j) =\frac{1}{\sqrt{A}}e^{\pm\frac{ i \theta_j}{2}}\cdot |\sin(\theta_j)|- e^{\mp\frac{ i \theta_g}{2}}\cdot |\sin (\theta_g)|
    \label{e2.1.6}
\end{equation}
After, substituting expression of dimensionless $A$,
\begin{equation}
    C(\theta_g, \theta_j) =\sqrt{\frac{M_j}{M_g}\left(\frac{1-\frac{\gamma}{\chi }\frac{Q_j}{M_j}}{1 \mp \frac{\gamma}{\chi}\frac{Q_g}{M_g}}\right)}e^{\pm\frac{ i \theta_j}{2}}\cdot |\sin(\theta_j)|- e^{\mp\frac{ i \theta_g}{2}}\cdot |\sin (\theta_g)|
\label{e2.1.7}
\end{equation}
from above equation (\ref{e2.1.7}) we express, that if \begin{equation}
1 > A >0 \begin{cases}
\text{and if $Q_g$ is \textit{positive}, then}
\begin{cases}
\text{both $\chi M_j > \gamma Q_j$ \& $\chi M_g > \gamma Q_g$ are true}\\
\text{both $\chi M_j < \gamma Q_j$ \& $\chi M_g < \gamma Q_g$ are true}\\
 \end{cases}\\
\text{if $Q_g$ is \textit{negative}, then }
\text{ $\chi M_j > \gamma Q_j$ must be true}\\
\end{cases}
\label{e2.1.8}
\end{equation}

It is interesting to find that irrespective to the charge carried by $Q_g$ , every case has the equal probability\footnotemark{\footnotetext{If A were found to be equal to zero then, $\frac{1}{\sqrt{A}}$ would be non-finite and if A were found be less than zero then the same expression would be Complex. }} of 0.5 of getting positive $A$.\footnotemark {\footnotetext{One might wonder how we are getting probability values. Let consider for positive S,then we can have two main cases depending on dual nature of |a|, $S_{n}=\frac{1+\frac{a}{b}}{1-\frac{c}{d}}>0$
and,
$S_{p}=\frac{1-\frac{a}{b}}{1-\frac{c}{d}}>0$ where, a,b,c and d are magnitudes hence positive, also S has to be finite, thus in each of the above cases we do get $d \neq c$, which is further getting subdivided into two cases as $d>c$ or $d<c$, now within each these two conditions we can independently have either $a>b$ or $a<b$. We see that in each cases there will be four sub-cases in respective sample space of $S_n$ and $S_p$, but only two are found to be favorable; i.e. when (d>c \& b>a) or (d>c \& b<a) otherwise negative, so 0.5 be the probability of $S_n>0$. Similarly when (d>c \& b>a) or (d<c \& b<a) otherwise negative, so 0.5 be the probability of $S_p>0$. Note that while $S_n>0$, relation between a and b is seem unimportant , thus only $d>c$ is crucial. }} However, despite of such equality , it seems that the nature prefers more towards having $Q_g<0$, Which itself says that for stable system central object should be positive, otherwise two like charges would be flying away from each other.  The reason to this argument is clearly visible, Notice that if the body $G_\pm$ happens to be a negative charged body of magnitude $Q_g$, then the nature do not have to carry any information regarding whether
$\chi M_g > \gamma Q_g$ or $\chi M_g < \gamma Q_g$, in addition to already carried information $\chi M_j > \gamma Q_j$. Such independence and freedom of choice push the system for having positively charged central body and the negatively charged body, which is placed outside. 
Moreover, we can also justify this claim mathematically\footnotemark{\footnotetext{Here, We continue with remarks which were made in just previous footnote. As we know that the sample spaces of S carries just four conditions, out of these only one is favourable if a is positive and two are favourable if a is negative, when $S>0$.  }}, Consider that for $A>0$ to be true, then probability of having only ($\chi M_j>\gamma Q_j$) oath to be true by 0.25, if $Q_g>0$ and 0.5, if $Q_g<0$. \textit{Hence, The nature gives preference to the system of two bodies having positive central core with negative charge enveloping the core when compared to positive-positive system of two bodies, preference is given by the factor of 2.} We also see that there is no chance of having $\chi M_j< \gamma Q_j$ , keeping $A>0$, as long as $Q_g$ is negatively charged.

\vspace{5pt}
\noindent
What is that C function (see, \ref{e2.1.6})? We know that $ 0^{\circ} \leq (\theta_g+\theta_j) <180^{\circ}$, but we do not have clue what this C function actually signifying?
Let consider the following case,
\begin{equation}
    C(\theta_g, \theta_j) =\frac{1}{\sqrt{A}}e^{\frac{ i \theta_j}{2}}\cdot (\sin(\theta_j))- e^{-\frac{ i \theta_g}{2}}\cdot (\sin (\theta_g))
    \label{e2.1.8b}
\end{equation}

\begin{figure}[H]
\centering
\begin{subfigure}{0.4\textwidth}
\includegraphics[width=0.9\linewidth, height=0.9\textwidth]{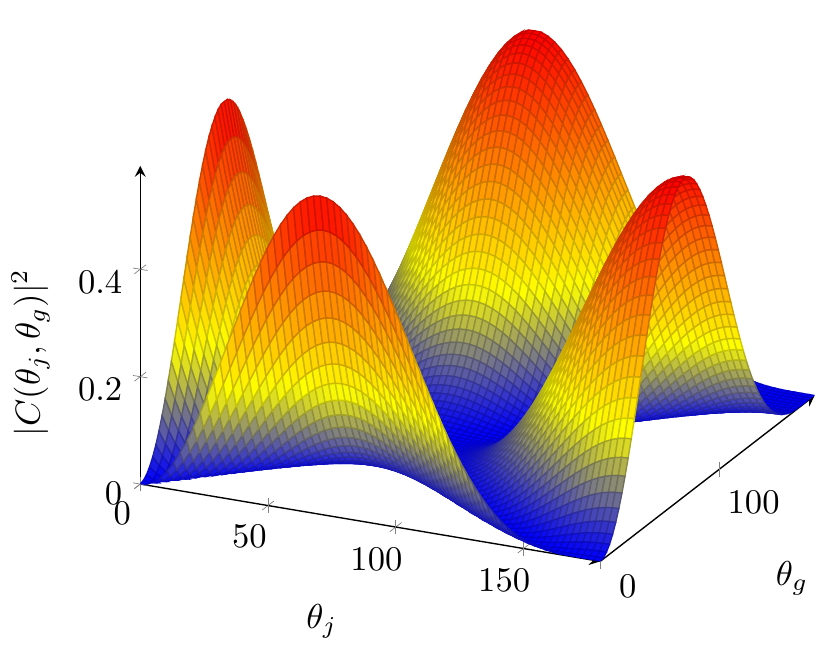}
\caption{}
\label{fig:c}
\end{subfigure}
\begin{subfigure}{0.4\textwidth}
\includegraphics[width=0.9\linewidth, height=0.8\textwidth] {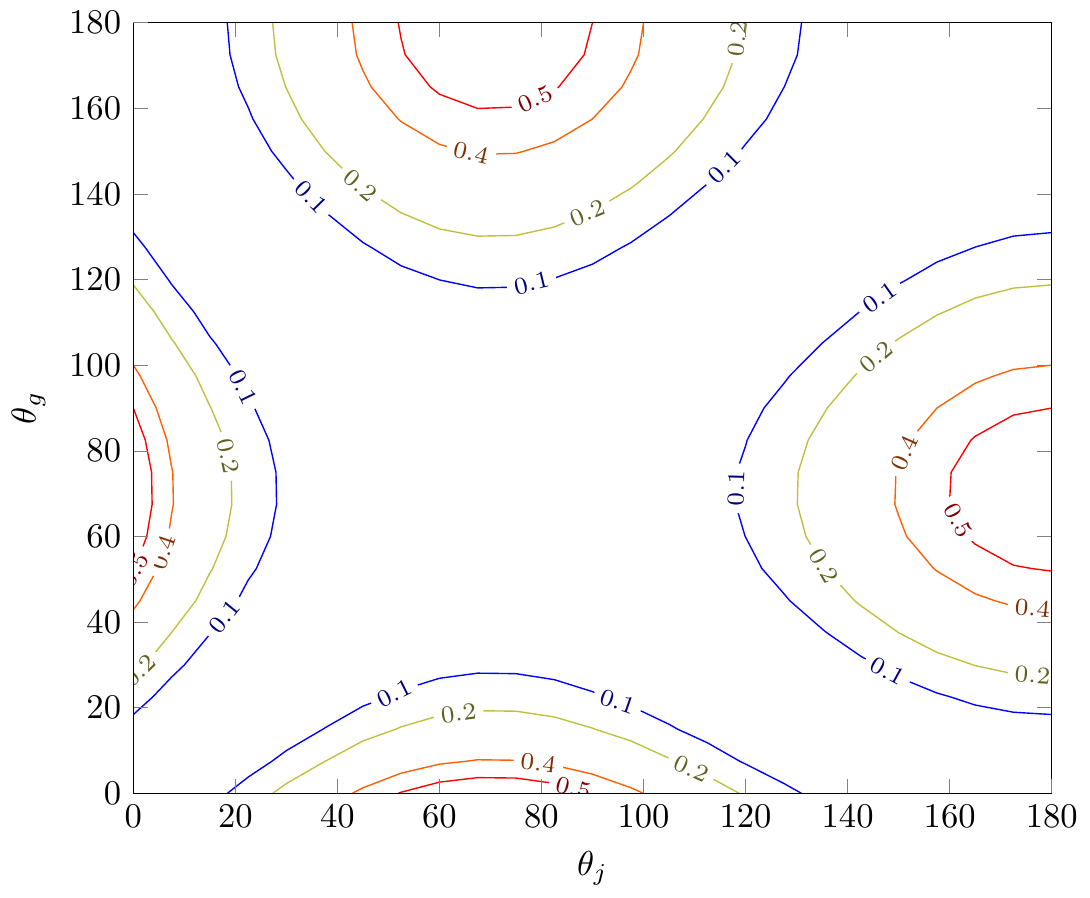}
\caption{}
\label{fig:c1}
\end{subfigure}

\caption{ \small The above figure (\ref{fig:c}) is showing the surface plot of the C-function squared,  $|C(\theta_g, \theta_j)|^2$; while keeping $\frac{1}{\sqrt{A}}\rightarrow1$ (see \ref{e2.1.8c}) and the contour plot in figure (\ref{fig:c1}).}
\label{fig:c3}
\end{figure}

If we multiply  $C(\theta_g, \theta_j)$ in above expression(\ref{e2.1.8b}) with its complex conjugate, $C(\theta_g, \theta_j)^*$ then we get a real function as,
\begin{equation}
\begin{split}
 |C(\theta_g, \theta_j)|^2 
 & =\left(\frac{1}{\sqrt{A}}\sin(\theta_j)\cos\left(\frac{\theta_j}{2}\right)-\sin(\theta_g)\cos\left(\frac{\theta_g}{2}\right)\right)^2\\
& +\left(\frac{1}{\sqrt{A}}\sin(\theta_j)\sin\left(\frac{\theta_j}{2}\right)-\sin(\theta_g)\sin\left(\frac{\theta_g}{2}\right)\right)^2
    \end{split} \label{e2.1.8c}
\end{equation}
  We should see how this expression is varying with $\theta_j$ and $\theta_g$ values in degrees, see figure(\ref{fig:c3}). 
   
  It seems that function repeats its values after certain period, as indicated in figure (\ref{fig:c}), four sided hood shaped curve. That is further supported by contour plot which is beautiful and diagonally symmetric to line passing through the origin. 
  Can we understand this C function more better? I think we should try to associated a meaning to equation (\ref{e2.1.6}) what is this function curving out in complex plane?
  
  For a complex function \citep{deaux2013introduction}
  
  $$Z(s)=A_{+}e^{is}+A_{-}e^{-is}$$ Where $A_{\pm}=\frac{a \pm b}{2} \in \mathbb{R} $ defines the parametric equation of a ellipse in the complex plane, provided $x(s)=a\cos(s)$ \& $y(s)=b\sin(s)$. Here, $x$,$y$ are the coordinates of any point on the ellipse; $a$, $b$ are the radius on the real and imaginary axes respectively and $s$ is the parameter, which ranges from $0$ to $2\pi$ radians.
 We can simply modify our C function as follow to turn it strongly similar to the above.
  \begin{equation}
  \begin{split}
    C
    & =\frac{\sin(\theta_j)}{\sqrt{A}}e^{\pm\frac{ i \theta_j}{2}}- \sin (\theta_g)e^{\mp\frac{ i \theta_g}{2}}\\
    & = K_{+}e^{\pm \frac{ i \theta_j}{2}} + K_{-}e^{\mp \frac{ i \theta_g}{2}}\\
     & = K_{+}e^{\pm is_g} + K_{-}e^{\mp is_j}\\
    \label{e2.1.8d}
     \end{split}
\end{equation}
  Where, $K_{+}=\frac{\sin(\theta_j)}{\sqrt{A}}$ and $K_{-}=-\sin(\theta_g)$, which can be made analogous to $A_{\pm}$ of $Z(s)$ in addition to it, $s_j=\frac{\theta_j}{2}$ and $s_g=\frac{ \theta_g}{2}$ are distinct and correspond to $s$ only if $|\theta_j|=|\theta_g|$, (see expression (\ref{s1r}) ) thence resulting to $K_+=-\frac{K_-}{\sqrt{A}}$ also $a=\sin(2s)\left(\frac{1}{\sqrt{A}}-1\right)$ and $b=\sin(2s)\left(\frac{1}{\sqrt{A}}+1\right)$, again only when, $s=\frac{\theta_j}{2}=\frac{\theta_g}{2}$. Generally, C function represents two sub-cases through itself,  say $C_+$ and $C_-$.
 \begin{equation*}
      C_{+} =K_{+}e^{is_j} + K_{-}e^{- is_g}
 \end{equation*}
 And,
  \begin{equation*}
      C_{-} =K_{+}e^{-is_j} + K_{-}e^{ is_g}
 \end{equation*}
 I expect when $\theta_j\neq\theta_g$ then it should be representing the non symmetric  horizontal and vertical ellipses for respective sub-cases of C function .

\subsubsection{Introducing \textit{quintessence } and reviewing the nature of \texorpdfstring{$\chi$} \textit{-coefficient} }
\label{s4}

We can also expand the above equation (\ref{e2.1.2}) in the terms of a generating function \citep{boas2006mathematical} for Legendre polynomials, 
\begin{equation}
\Phi(x,h)=\left(1-2xh+h^2\right)^{-1/2}=\sum_{l=0}^{\infty}{h^l P_l(x)}
\label{e2.1.9}
\end{equation}

Where $|h|<1$ and $ P_l(x)$ are the Legendre polynomials , provided that we have defined $ T \propto \frac{1}{\vec{|f|}}$, we call $T$ as \textit{quintessence }\footnotemark{\footnotetext{\textit{quintessence } here, it is expressing a meaning associated with reciprocal of the  magnitude of fields strength, generally in classical physics charge or mass is expressed as $(q,m...) =|\frac{Force}{\text{fields  strength}}|$, so one can relate to \textit{quintessence  },$T$ as a collective term.}}. Note initially we have taken $\left|\frac{\chi G_g\mp E_g}{\chi G_j-E_j}\right|<1$, which is indeed analogous to $|h|$ and $x$ is to $cos(\theta)$ in the generating function.
then,
\begin{equation}
T\propto \frac{1}{|\chi G_j-E_j|}\left[1-2\left(\frac{\chi G_g\mp E_g}{\chi G_j-E_j}\right)\cos(\theta)+\left(\frac{ \chi G_g\mp E_g }{\chi G_j-E_j}\right)^2\right]^{-1/2}
\nonumber
\end{equation}
using equation (\ref{e2.1.9}) we expand the above expression as
\begin{equation}
T\propto \sum_{l=0}^{\infty} \left(\frac{\chi G_g\mp E_g}{\chi G_j-E_j}\right)^l\frac{P_l(\cos\theta)}{|\chi G_j-E_j|}
\nonumber
\end{equation}
The factor, $\left(\frac{\chi G_g\mp E_g}{\chi G_j-E_j}\right)^l=\left[\frac{l_j^2}{l_g^2}\left(\frac{\chi G_NM_{g}\mp \frac{1}{4\pi\epsilon_0}Q_g}{\chi G_NM_{j}-\frac{1}{4\pi\epsilon_0}Q_j}\right)\right]^l=\left[\frac{l_j^2}{l_g^2}(A)\right]^l=e^{\pm i\theta l}$ using above relations. Then, 
\begin{equation}
T\propto \sum_{l=0}^{\infty} \frac{P_l(\cos\theta)}{|\chi G_j-E_j|}e^{\pm i\theta l}
\label{e2.1.10}
\end{equation}
Note, $T$ of body $G_{\pm}$ is independent of its own mass $M_g $ and  electric charge $ Q_g$. If demanded, one should more correctly denote  $T$ as $T^g$, Which would reminds us of \textit{quintessence } of object $G_{\pm}$ as $T^g$, however, currently we will not be doing such things. As for now, looking back to expression (\ref{e2.1.10}), from where we take out first few terms from the summation series as,

\begin{equation}
T\propto \frac{1}{|\chi G_j-E_j|}+\frac{\cos\theta}{|\chi G_j-E_j|}e^{\pm i\theta}+\frac{1}{2}\frac{3(cos\theta)^2 -1}{|\chi G_j-E_j|}e^{\pm 2 i\theta}+...
\label{e2.1.11}
\end{equation}
Since,\\
$P_0(cos\theta)=1  $\\
$P_1(cos\theta)=cos(\theta) $\\
$P_2(cos\theta)=\frac{1}{2}\left(3(cos\theta)^2 -1\right) $... and so on.\\
Looking at the first term\footnotemark{\footnotetext{ Note that $T_0$ is only term in entire series which is real and probably scalar. }} to the right side in equation (\ref{e2.1.11}) as $T_0 \propto \frac{1}{|\chi G_j-E_j|}$
we infer that
$T_0$  is the real quantity, but further terms of $T$ i.e. $T_1$, $T_2$... are oath to be complex in nature.
This is causing a difficulty in assigning a meaning to those terms. Therefore let rewrite the above equation (\ref{e2.1.10}) as follows,
\begin{equation}
T\propto \sum_{l=0}^{\infty} \frac{P_l(\cos\theta)}{\left|\frac{\chi G_j}{e^{\pm i\theta l}}-\frac{E_j}{e^{\pm i\theta l}}\right|}
\label{e2.1.11b}
\end{equation}
 see here, we can define complex quantities namely as hyper-true mass and hyper-true charge respectively as $D_l^j=\frac{M_j}{e^{\pm i\theta l}}$ and $I_l^j=\frac{Q_j}{e^{\pm i\theta l}}$.
 then, 
 \begin{equation}
T\propto \frac{l_j^2}{G_N}\sum_{l=0}^{\infty} \frac{P_l(\cos\theta)}{\left|\chi D_l^j-\gamma I_l^j \right|}
\label{e2.1.11c}
\end{equation}
Interestingly, if we say for a while $T_0=0$, then we get following condition, 
\begin{equation}
    \chi M_j >\gamma Q_j
    \label{e2.1.12}
\end{equation}
The above relation directly advocates to the statement made about nature of A in above subsection (\ref{section2.1.3}) that $Q_g<0$ (if it happens to exist), hence if the nature ever try to pair up two bodies then she should prefer by pairing up positive and negative charged bodies. Which is consistent with the argument after observing the equation (\ref{e2.1.8}).
Let me comment few things about \textit{quintessence }; I strongly think that mass and charge should get encapsulated under common [...], making us to get familiar with the profound truth. If we consider that \textit{quintessence } might be a real and scalar quantity, then the above expansion of $T$ might be thought analogous to multipole expansion of electric potential $V$, $$V(\mathbf {r} )=V_{\text{mon}}(\mathbf {r} )+V_{\text{dip}}(\mathbf {r} )+V_{\text{quad}}(\mathbf {r} )+\ldots \label{ee1}$$
 analogously, we should also have, for \textit{quintessence } $T$, 
  $$T \propto T_{0}+T_{1}+T_{2}+\ldots \label{ee2}$$
  
  Additionally, we know that each ascending term in potential's multipole expansion is respectively constituting \textit{electric moment} as tensor of rank 0, 1, 2, and so on, i.e.  in  $V_{\text{mon}}=\dfrac{q}{ 4\pi \epsilon _0 r}$, Here $q$ is tensor of rank zero (scalar) or electric monopole moment. However, when we look back to our $T$, we notice, the trend of ranking of tensor in $V$ is exchanged by the \textit{type of number} in case of $T$. More clearly , $T_0 \propto \frac{1}{|\chi G_j-E_j|}$, Here denominator is a complex number with imaginary part zero i.e. a real number but for $T_{n>0}$ their denominators or hyper true quantities are indeed complex numbers. If we could sense the underlying idea that in \textit{quintessence } what complex number does seem analogous to what tensor is doing in $V$. Fortunately, we have mathematical tools ( for rank one tensor, dot product) which maps tensor of rank greater than equal to one to a scalar, as a result $V$ always remains a real and scalar quantity.
  
  However, only speculation can be made here for the case of \textit{quintessence }, which requires some kind of mathematical tool (convert complex hyper quantities to real ones) or physical interpretation such that we get $T$ as Real physical quantity along with being scalar.
  
  Instead of focusing on all these thing, we could have alternative resolution for expression (\ref{e2.1.11c}), if we just assume that,  $\theta=0$ as one of the result.\footnotemark{\footnotetext{One might think that we have used expression(\ref{s1r}), Since that was obtained by making  $|\vec{f}|=0$, and here we  also know T is defined as inversely proportional to net field; \textit{did i make a mistake}? I feel that i might  have not done any mistake. As we know that T has its roots in expression (\ref{e2.1.2}) and making $\theta=0$ would turn out $x=1$ analogously in the above generating function. Since, generating function does not mention about any condition on x, see equation-($5.1$) in page-569 of reference \cite{boas2006mathematical}. Thus our statement on $\theta$ is independent of result obtained as (\ref{s1r}).
  }}
  We get, using property of Legendre polynomial $P_l(1)=1$.
  \begin{equation}
  \begin{split}
      T & \propto \frac{l_j^2}{G_N}\sum_{l=0}^{\infty} \frac{P_l(\cos(0))}{\left|\chi D_l^j-\gamma I_l^j \right|}\\
      & \propto  \frac{l_j^2}{G_N}\sum_{l=0}^{\infty} \frac{P_l(1)}{\left|\chi \frac{M_j}{e^{0}}-\gamma \frac{Q_j}{e^{0}} \right|}\\
        & \propto  \frac{l_j^2}{G_N}\sum_{l=0}^{\infty} \frac{1}{\left|\chi M_j -\gamma Q_j\right|} 
        \label{e2.1.13a}
  \end{split}
\end{equation}

Ultimately, we see that above equation is independent of $l$  and thus we should remove summation sign and we get,
\begin{equation}
     T \propto  \frac{l_j^2}{G_N\left|\chi M_j -\gamma Q_j\right|}\label{e2.1.13b}
\end{equation}

Thus, we may comment that \textit{quintessence } is turned out to a real physical quantity under consideration of assumed result $\theta=0$ , note that dimension of the same is \textit{proportional} to $[M^{-1} L^{-1} T^{3} A]$.
Let for another source object $Q$, we can have associative \textit{quintessence } for test object $P$, say $T^p$,
then from above proportionality relation, we get,
\begin{equation}
\frac{T^g}{T^p} =\frac{\left|\chi_q M_q -\gamma Q_q\right|}{\left|\chi_j M_j -\gamma Q_j\right|} \frac{l^2_j}{l^2_q}
    \label{e2.1.13c}
\end{equation}
Since, we are currently not  aware of whether Chi constant is universal or not, thus we have used sub-scripted $\chi$ in above equation.

  \vspace{5pt}
  \noindent
Now, we should look into the behaviour that $\chi$ shows.
Since we have been knowing the Permittivity of free space, $\epsilon_0 \approx 8.8541878128\times 10^{-12}\frac{\text{C}^2 {\cdot} \text{s}^2}{\text{kg} {\cdot} \text{m}^3}$ and universal gravitational constant $G_N=6.67430 \times 10^{-11} \frac{\text{m}^3}{\text{kg} {\cdot}\text{s}^2}$. Now, we know that $\gamma=\frac{1}{4\pi \epsilon_0 G_N}$, which is equal to $1.3465909222\times 10^{20} \frac{\text{kg}^2}{\text{C}^2}$. If one is interested of having the value of $\gamma$ in terms of $\frac{\text{u}^2}{\text{e}^2}$ where u is unified mass, $1\text{kg} =6.0221407621\times10^{26} \text{u}$
and e is the elementary charge, $1C\approx 6.24150907446\times10^{18} \text{e}$.

Substituting these conversion factors, gives the values of  $$\gamma=1.253597955\times10^{36}\frac{\text{u}^2}{\text{e}^2}$$ 

 Now, we will try to understand whether Chi coefficient, $\chi=\frac{\gamma Q_j}{M_j}$  from equation (\ref{e2.1.12}) is constant or not and then accordingly, we will review our assumption about it.
 We will consider two cases, first of it would consider object $J_+$ as a nuclei and in second case we take object $J_+$ to be a electron\footnotemark{\footnotetext{Note we are  momentarily breaking one declaration that $Q_j$ should be positive charged object but electrons are obviously not positive; So one may consider them to be positrons, since both have same mass.}}. Then, we  plot the graph for most of the isotopes of atoms.

 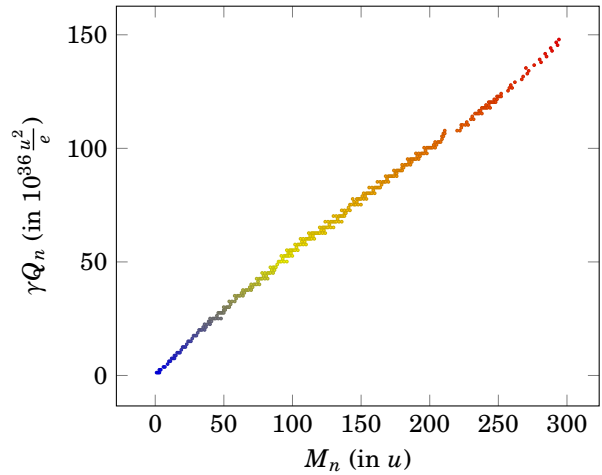
\begin{figure}[H]
   \centering
   \begin{tikzpicture}
    \begin{axis}
       [xlabel={$M_{n}$ (in $u$)},ylabel={$\gamma Q_n$ (in  $10^{36}\frac{u^2}{e})$}],
       \addplot[only marks,
    scatter,
    mark size=0.5pt
    ]table{tizk/nm.txt};
    \end{axis}
   \end{tikzpicture}
    \caption{\small  The body $J_+$ with nuclear masses $M_j$ in unit $u$ is taken along in x axis for 350+ different nucleus including with their isotopes in step-wise increasing atomic numbers (Nuclear charges are marked by their Z numbers), and positive nuclear charge $Q_j$ is in $e$ units times $\gamma$ is taken along y axis. Its slope seems to be constant.}
    \label{fig:8}
    \end{figure}

    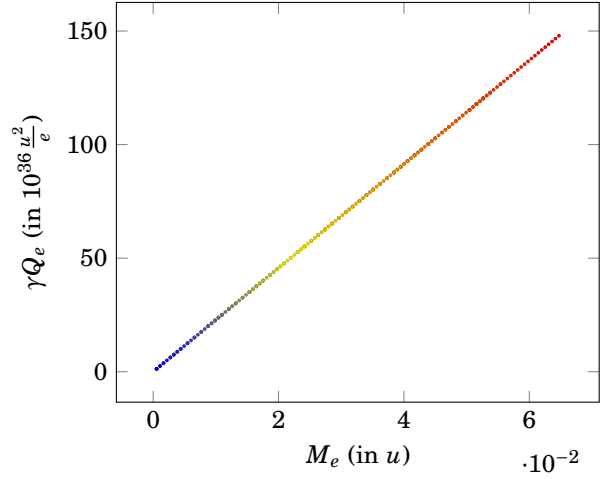
\begin{figure}[H]
   \centering
   \begin{tikzpicture}
    \begin{axis}
       [xlabel={$M_{e}$ (in $u$)},ylabel={$\gamma Q_e$ (in  $10^{36}\frac{u^2}{e})$}],
       \addplot[only marks,
    scatter,
    mark size=0.5pt
    ] table{tizk/em.txt};
    \end{axis}
   \end{tikzpicture}
    \caption{\small Here, We have considered the $M_j$ to be non-nuclear mass (total electrons mass) and electron's charge as $Q_j$. Such linear dependence is trivial, as isotopes have same Z numbers}
    \label{fig:9}
\end{figure}

\begin{figure}[H]
   \centering
   \begin{tikzpicture}
    \begin{axis}
       [xlabel={Z},ylabel={\small $\frac{M_n}{M_e}$ }],
       \addplot[only marks,
    scatter,
    mark size=0.5pt
    ] table{tizk/ratio.txt};
    \end{axis}
   \end{tikzpicture}
    \caption{\small Note that mean values of y axis would lies within 5000 to 4000 range or around 4500  and the number 4512.44076 falls exactly within that range.}
    \label{fig:10}
\end{figure}
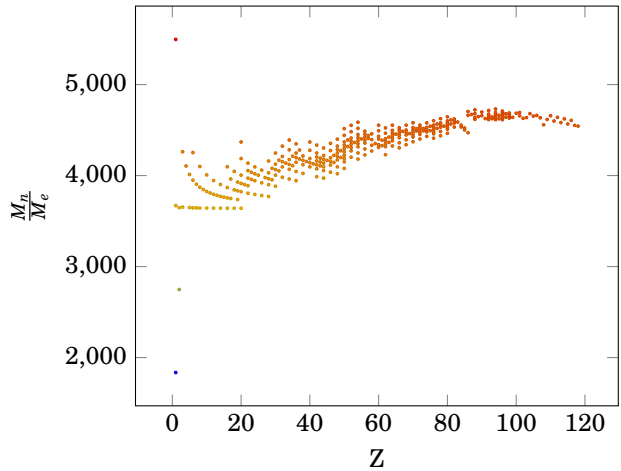

By above plots we see that slope of the plot (\ref{fig:8}) is $0.5064\times 10^{36} \small{\frac{\text{u}}{\text{e}}}$  and graph (\ref{fig:9}) is $2285.1 \times 10^{36} \small{\frac{\text{u}}{\text{e}}}$. Taking the ratio of these two slope values as $4512.44076$. One may confirm it by the plot (\ref{fig:10}), which is showing the graph between the ratio nuclear mass to non-nuclear mass verses atomic number Z. 

 We see that our assumption about $\chi$ factor is indeed promising and correct. However, we understand the $\chi=\gamma\frac{\Delta Q_j}{\Delta M_j}$ found to be different on nuclear and non-nuclear scales (see figures \ref{fig:8} \& \ref{fig:9} ), thus it may not be an universal constant. Therefore, sub-scripted $\chi$ is a valid in expression (\ref{e2.1.13c}). Since this Chi factor is highly sensitive to order of size of the object we have considered, it may act as a indicator, which could always hint us about the scale of objects, we are working with. We know a nucleus is both more massive and larger in size than a electron indicating that electron, a particle should be denser than nucleus. We recently seen that $\chi_e > \chi_n$ by the the order of $10^3$. Which is proportional to density of  body and inversely to mass of the body. Thus $\chi$  seem to be inversely proportion to the dimension of the body. 
 
 What if we try to work on unit scale, therefore
 Let say $\chi_0=1\small{\frac{\text{kg}}{\text{C}}}$
 then we find,
 \begin{equation*}
 \frac{\Delta M_j}{\Delta Q_j}=\frac{\gamma}{\chi}=1.3465909222\times 10^{20} \small{\frac{\text{kg}}{\text{C}}}
 \end{equation*}
 Or,
 A body K is said to lie in \textit{unit scale}, if its 1 C of electric charge is being carried by $1.346\times 10^{20} $ kg of its mass\footnotemark{\footnotetext{Another way for stating this, $1$ kg of mass if carries about $0.0742\times 10^{-19} $ C of electric charges. Strangely, one can consequently determine that an electron should be composed of at least $10^{31}$ further elementary bodies.}}; then any object larger than K in volume should be regarded as a large body and any body smaller than that is need to be called as small body.
 I suggest that, We should do intense works to discover whether universe distinguish between scales?
 
\section{Conclusion}
In this study we tried to understand the statics of two non-neutral objects. While  formulation We did not ignore the effect of gravitational influences in presences of electrostatics fields strength. We learned something new about how nature always desires to create structures in which objects of positive properties are to be present at the center. We learned about \textit{quintessence} which is a quantity whose direct concern is with its own mass and value of electric charge; however it was found to be depending inversely on linear combination of some parameters of the source body (in our case it was the body $J_+$). It is certain that when \textit{ $a\rightarrow1$}, allowing us to have neutral plane, where net field is zero, Which later helps us to associate meaning to so called C-function as equation of ellipse in complex plane and existence of \textit{specific field parabolas} are also getting revealed. We felt the necessity of  having physical interpretation of C function, which mathematically resembling equation of ellipse in complex plane. For non-zero net field strength, \textit{quintessence } was turned out to be most significant outcome which 
enabled us to visualize the object's defining parameters such as mass and electric charges through single means. Lastly, We did verify the constancy of a factor $\chi$ almost constant but it is highly sensitive to order of size of the object under study. This study is currently incomplete and we hope that in the future we will try to understand the dynamics of two objects that are full of charge. Right now we should not rush to come to concrete conclusions, that is my opinion.
\section{Acknowledgement}
I take immense pleasure in acknowledging Dr. Sharad K. Yadav, Assistant professor, Department of Physics, Sardar Vallabhbhai National Institute of Technology, Surat for his kind guidance during January 2022 and May 2022.
\bibliography{refs}

\end{document}